\def\half{\frac{1}{2}}
\def\nll{\hfil\linebreak\noindent} 
\def\hi#1#2{$#1$\kern -2pt-#2} 
\def\hy#1#2{#1-\kern -2pt$#2$} 
\def\dbox#1{\hbox{\vrule 
\vbox{\hrule \vskip #1\hbox{\hskip #1\vbox{\hsize=#1}\hskip #1}\vskip #1 
\hrule}\vrule}}  
\def\qed{\begin{flushright}~{\dbox{0.05true in}}\end{flushright}} 
\def\heading#1{\hfil\linebreak\noindent{\bf #1}~~}
\def\third{{\scriptstyle {1\over 3}}}
\documentclass{iopart}
\usepackage{graphicx}
\begin{document}
\hspace*{2.5 in}CUQM-120, HEPHY-PUB 836/07\\


\vspace*{0.4 in}

\title{Binding energy of semirelativistic $N$-boson systems}
\author{Richard L. Hall}
\address{Department of Mathematics and Statistics, Concordia University,
1455 de Maisonneuve Boulevard West, Montr\'eal,
Qu\'ebec, Canada H3G 1M8}
\author{Wolfgang Lucha}
\address{Institute for High Energy Physics, Austrian Academy of Sciences,
Nikolsdorfergasse 18, A-1050 Vienna, Austria}
\eads{\mailto {rhall@mathstat.concordia.ca}, \mailto {wolfgang.lucha@oeaw.ac.at}}
\begin{abstract}General analytic energy bounds are derived for $N$-boson systems governed by
semirelativistic Hamiltonians of the~form 
$$H=\sum_{i=1}^N({\bf p}_i^2+m^2)^{1/2}+\sum_{1=i<j}^NV(r_{ij}),$$
 where $V(r)$ is a static attractive pair potential.
A translation-invariant model Hamiltonian $H_c$ is constructed.  We conjecture that $\langle H\rangle \ge \langle H_c\rangle$ generally, and we prove this for $N = 3$, and for $N= 4$ when $m=0.$ The conjecture is also valid generally for the harmonic oscillator and in the nonrelativistic large-$m$ limit.  This formulation allows reductions to scaled 3- or 4-body problems, whose spectral bottoms provide energy lower bounds.  The example of the ultrarelativistic linear potential is studied in detail and explicit upper- and lower-bound formulas are derived and compared with earlier bounds.
\end{abstract}
\pacs{03.65.Ge, 03.65.Pm\hfil\break
{\it Keywords\/}:
Semirelativistic Hamiltonians, Salpeter Hamiltonians, boson systems}
\vskip0.2in
\maketitle
\section{Introduction}
One-body Hamiltonians $H$
composed of the relativistic expression $\sqrt{{\bf p}^2+m^2}$ for the
kinetic energy of particles of mass $m$ and momentum ${\mathbf p}$ and of a
coordinate-dependent static interaction potential $V({\bf r}),$ defined as
operator sum
$$H=\sqrt{{\bf p}^2+m^2}+V({\mathbf r}),$$
provide a simple but very
efficient tool for the description of relativistically
moving particles~\cite{BSE,SE,Lucha:O}.
 They have been used, for
instance, for the description of hadrons as bound states of quarks
\cite{Lucha91}. One of the advantages of this kind of
semirelativistic treatment is that its generalization to the many-body
problem is straightforward to formulate \cite{Lieb96}. A semirelativistic Hamiltonian for a
system of $N$ identical particles interacting by pair potentials $V(r_{ij})$
is given by
$$H=\sum_{i=1}^N\sqrt{p_i^2+m^2}+\sum_{1=i<j}^NV(r_{ij}).\eqno{(1.1)}$$
We use the notational simplification
$p\equiv\|{\bf p}\|,$ $r\equiv\|{\mathbf r}\|$, or $r_{ij}\equiv\|{\mathbf r}_i-{\mathbf
r}_j\|,$ whenever no ambiguity is introduced by so doing. Many approaches to
such many-body problems for identical particles employ the very powerful
constraint of permutation symmetry to generate their reduction to a two-body
problem with a Hamiltonian ${\mathcal H}$ whose spectrum is used to
approximate the many-body energy eigenvalues or to generate a lower energy
bound. This reduction may be effected in various ways, which leads to the
problem of finding the most effective reduced problem, the one which would
provide the {\it highest\/} lower bound. In one analysis \cite{Hall67a} involving pseudo-fermions (where the necessary permutation antisymmetry is carried entirely by the spatial part of the wave function), an optimization is considered over a real parameter which characterizes the degree of orthogonality of the matrix $B$ that defines the relative coordinates. For boson systems, an orthogonal $B$ is best possible; one such choice is the Jacobi coordinate system that we shall use in Section~2 below. 

For the boson problem, perhaps the most immediate reduction is what we have called the simple or $N/2$ bound based on the equality $\langle H\rangle = \langle H_2 \rangle,$ where
$$H_2 = {N\over 2}\left[\sqrt{p_1^2+m^2}+ \sqrt{p_2^2+m^2} + (N-1)V(r_{12})\right].\eqno{(1.2)}$$
The $N/2$ bound is then the bottom $E_2$ of the spectrum of the scaled two-body Hamiltonian $H_2.$ We have explicitly for this bound
$$E \geq E^L_{N/2} = N\inf_{\psi}\left(\psi, \left[\left(p^2 + m^2\right)^{\half} + {{N-1}\over{2}}V(r)\right]\psi\right).\eqno{(1.3)}$$
If this reasoning is applied to the Schr\"odinger harmonic-oscillator problem, one finds for large-$N$ that $E_{N/2}^L \rightarrow E/\sqrt{2},$ whereas a reduction based on Jacobi coordinates \cite{Hall67} yields $E_L = E.$  We note parenthetically that the $N/2$ bound is equivalent to using a non-orthogonal coordinate system consisting of a centre-of-mass coordinate and $N-1$ pair distances \cite{Hall74}. Similarly, one obtains dramatic improvement over the $N/2$ lower bound  if Jacobi coordinates are used for the Salpeter harmonic-oscillator problem \cite{Hall03}. We have obtained improved lower bounds for potentials which are  convex transformations $V(r) = g(r^2)$ of the oscillator \cite{Hall04}, and also, by very special reasoning, for the
 gravitational potential \cite{Hall06}, $V(r)=-v/r,$ $v>0.$ In the present paper
we look for good lower bounds that are valid for general attractive potentials,
for example, of the form $V(r) = -v/r + b r,$ $v \geq 0, b > 0.$

Since the spectrum of the semirelativistic many-body Hamiltonian $H$ can be characterized variationally, it is straightforward to
find upper energy bounds with the aid of a suitable trial function.  The principal difficulty is to find a good general lower bound. We achieve this for $N = 3,$ and for the case $m = 0, N = 4.$ These partial results then
allow the construction of corresponding lower bounds based on reductions of the many-body problem respectively to scaled $N =3$ and $N=4$
systems.  A formulation that unifies these results and all the known earlier partial results may be expressed as a lower-bound conjecture, which then becomes a theorem for each case that is proved. 

\section{Lower-bound conjecture}
We first consider a model $N$-body Hamiltonian. This model has been constructed so that it essentially yields the corresponding nonrelativistic result in the limit $m\rightarrow\infty.$ We are guided in the first instance by the centre-of-mass identity and inequality~\cite{Hall67}
$$\sum\limits_{i = 1}^{N}{\mathbf p}_i^2 = {1\over N}\sum\limits_{1=i<j}^{N}({\mathbf p}_i -{\mathbf p}_j)^2 + \frac{1}{N}\left(\sum\limits_{i}^{N}{\mathbf p}_i\right)^2\ge{1\over N}\sum\limits_{1=i<j}^{N}({\mathbf p}_i -{\mathbf p}_j)^2.\eqno{(2.1a)}$$
For the corresponding semirelativistic problem, we lose this transparent algebraic inequality and must instead recover whatever can be proved to be true on the average.  In a nutshell, this is the technical difficulty we must face in this paper.  The Schr\"odinger $N$-body Hamiltonian $H_S$ with the centre-of-mass kinetic energy removed and $\hbar = 1$ is therefore given by
$$H_S = \sum\limits_{1=i<j}^{N}\left[\frac{1}{2mN}({\mathbf p}_i -{\mathbf p}_j)^2 +V(r_{ij})\right].\eqno{(2.2)}$$
In Jacobi coordinates $[\rho] = B[{\mathbf r}],$ where $B$ is an orthogonal $N\times N$ matrix with first row having entries all equal to $1/\sqrt{N},$ $\rho_2 = ({\mathbf r}_1- {\mathbf r}_2)/\sqrt{2},$ and conjugate momenta $[\pi] = (B^t)^{-1}[{\mathbf p}] = B[{\mathbf p}],$ the equality in (2.1a) may be re-written simply
$$\sum\limits_{i = 1}^{N}{\mathbf p}_i^2 = \pi_1^2 + \sum\limits_{i = 2}^{N}\pi_i^2.\eqno{(2.1b)}$$
Meanwhile, if $\Psi(\rho_2,\rho_3,\dots,\rho_N)$ is a normalized translation-invariant $N$-boson wave function, we have \cite[Eqs.~(6) and (7)]{Hall06}: 
$$\left(\Psi, \pi_i^2\Psi\right) = \left(\Psi, \pi_2^2\Psi\right),\quad \left(\Psi, \rho_i^2\Psi\right) = \left(\Psi, \rho_2^2\Psi\right),\quad i = 2,3,\dots.\eqno{(2.3)}$$
We note parenthetically, for future reference, that with Jacobi coordinates we have the following explicit expression for ${\mathbf p}_N$:
$${\mathbf p}_N = \frac{\pi_1}{\sqrt{N}} - \sqrt{\frac{N-1}{N}}\pi_N.\eqno{(2.4)}$$

When either the kinetic energy is a quadratic expression, as for all Schr\"odinger problems \cite{Hall67}, or if the potential $V(r)$ is the harmonic oscillator $V(r) = kr^2$ \cite{Hall04}, then these relations play a key role in the construction of a lower-bound model.  Our purpose here is to make a reduction for the Salpeter problem and general $V(r)$, that is for problems for which neither the kinetic energy nor the potential energy has a simple quadratic form.  We focus our attention on the kinetic energy since any progress made here would be potential independent.  With these goals, the model $N$-body Hamiltonian we have constructed is given by
$$H_c= \sum_{1=i<j}^N\left[\sqrt{\gamma^{-1}({\mathbf p}_i-{\mathbf p}_j)^2 + \left({{2m}\over{N-1}}\right)^2}~+~ V(r_{ij})\right]\eqno{(2.5a)}$$
or, equivalently,
$$H_c= \sum_{1=i<j}^N\left[{\gamma}^{-1}\sqrt{\gamma({\mathbf p}_i-{\mathbf p}_j)^2+ (mN)^2}~+~ V(r_{ij}) \right],\eqno{(2.5b)}$$
where $\gamma = {N\choose 2} = \half N(N-1)$ is the binomial coefficient. In the Schr\"odinger limit $m\rightarrow \infty,$ we find  $H_c \rightarrow mN + H_S,$ where $H_S$ is exactly the corresponding Schr\"odinger $N$-body Hamiltonian with the centre-of-mass kinetic energy removed, given in (2.2).  Meanwhile, for the special case $N = 2$ of the semirelativistic problem itself we recover the well-known $2$-body Salpeter Hamiltonian
$$H = 2\sqrt{\left({{{\mathbf p}_1-{\mathbf p}_2}\over 2}\right)^2 + m^2}~~+~~V(r_{12}).\eqno{(2.6)}$$
If we use new conjugate coordinates, we may write $r = \|{\mathbf r}\| = r_{12}$ and $p = \|{\mathbf p}\| = \|({\mathbf p}_1-{\mathbf p}_2)/2\|,$ and then we have from (2.6)
$$H = 2\sqrt{p^2+m^2}~~+~~V(r).\eqno{(2.7)}$$
The idea is eventually to obtain an $N$-body lower bound which is the bottom of the spectrum of a scaled version of (2.6), namely
$${\cal H} = \beta\sqrt{\lambda p^2+ m^2}~~+~\gamma V(r),\quad \beta,\lambda, \gamma > 0.\eqno{(2.8)}$$
Meanwhile, the Salpeter Hamiltonian $H$ itself is given by (1.1). We now suppose that $\Psi$ is a translation-invariant normalized boson trial function. We consider expectations with respect to $\Psi$ and we first observe that the permutation symmetry of $\Psi$ implies the equality
$$\langle H_c\rangle = \langle {\cal H}\rangle,\quad {\rm where}\quad \beta = N,~~ \lambda = {{2(N-1)}\over{N}},~~\gamma = \half N(N-1).\eqno{(2.9)}$$\medskip
With these explicit values for the parameters $\{\beta,\lambda,\gamma\}$ in ${\mathcal H,}$ we are now able to formulate the central idea of this paper explicitly, namely

\heading{Conjecture}
$$\langle H\rangle \geq \langle {\mathcal H}\rangle .\eqno{(2.10)}$$\medskip
\noindent This implies the following explicit conjectured lower energy bound
$$E \geq E_{c}^L = N\inf_{\psi}\left(\psi, \left[\left({{2(N-1)}\over{N}}p^2 + m^2\right)^{\half} + {{N-1}\over{2}}V(r)\right]\psi\right).\eqno{(2.11)}$$
We can recover all earlier sharp bounds from this expression.  We immediately recover the Schr\"odinger bounds \cite{Hall67} in the $m\rightarrow\infty$ limit (2.5).  If we now assume (2.11) is true as it stands for $m \ge 0,$ and $V(r) = v r^2,$ we recover our earlier bounds \cite{Hall03} for the semirelativistic oscillator.  For $m >0,$ and $V(r) = - v/r,$ we recover our earlier sharp bounds for the gravitational problem \cite{Hall06}.  Meanwhile, the bounds we prove in the present paper establish a wider range of validity for this conjecture.  For example, our Theorem~3 below establishes (2.11) for $m \ge 0$ and $N = 3$ in dimension $d = 3;$ and Theorem~4 establishes the case $m = 0,\  N= 4.$ At present we know of no counter example.

If we compare (2.5b) with (1.1) we see that the expectation of the difference may be written
$$\langle H-H_c\rangle =\langle H-{\mathcal H}\rangle  = \langle \delta(m, N)\rangle,\eqno{(2.12)}$$
where
$$
\delta(m, N) = \sum_{i = 1}^{N}\sqrt{{\mathbf p}_i^2 + m^2}\  -\  {{2}\over{N-1}}\sum_{1=i<j}^{N}
 \sqrt{{{N-1}\over{2N}}({\mathbf p}_i -{\mathbf p}_j)^2+ m^2}.\eqno{(2.13)}
$$

All our lower-bound results follow from the positivity (strictly speaking, non-negativity) of $\langle\delta(m, N\rangle,$ when this can be established. We consider immediately the case $\{m = 0,\ N = 2\}$: the kind of reasoning we use in this case is generalized for the other cases. The approach we adopt is to think of the mean-value computation in momentum space where the momentum vectors ${\mathbf p}_i$ are multiplicative operators: these vectors form geometrical figures whose edges are the corresponding norms $\|{\mathbf p}_i\|;$ mean values $\langle\|{\mathbf p}_i\|\rangle = d$ are considered last.  For example, with $N = 2,$ the three vectors $\{{\mathbf p}_1,\  {\mathbf p}_2,\ {\mathbf p}_1-{\mathbf p}_2\}$ form the sides of a triangle. The observation that, as a consequence of the triangle inequality and boson symmetry, the largest possible value for $\langle\|{\mathbf p}_1 - {\mathbf p}_2\|\rangle$ is $2d,$ then establishes positivity in this case.  For $m > 0$ the argument must be adjusted accordingly. We shall consider this point in more detail in Section~5 below, for the more interesting case $N = 3$ and $m > 0.$  In order to prepare for what might be called `stochastic geometry', we consider first $N = 3$ and $m = 0$, although this is a special case of the more general problem $m \ge 0$ discussed later. As we have remarked above, for the corresponding Schr\"odinger problem for general $V(r),$ or for the Salpeter harmonic-oscillator problem with $V(r) = k r^2,$ a quadratic form is involved either in the kinetic- or the potential-energy term: for both of these problems, the conjecture follows as a result of the general quadratic mean-value identities (2.3) in Jacobi coordinates. For the Salpeter problems with general $V$, which is the subject of the present paper, the quadratic expressions (in momentum space) always appear inside the square-root sign, so these identities do not immediately apply. The general inequality $\langle \|{\mathbf p}\|\rangle \le  \langle \|{\mathbf p}\|^2\rangle^{\half}$ does not remove this difficulty. 

\section{Proof in the case $m = 0$,~~$N = 3$} 
We have the following definition from (2.13):
$$\delta(0, 3) 
= \|{\mathbf p}_1\| + \|{\mathbf p}_2\|+ \|{\mathbf p}_3\|-{1\over{\sqrt{3}}}\left(\|{\mathbf p}_1 - {\mathbf p}_2\|+\|{\mathbf p}_1 - {\mathbf p}_3\|+\|{\mathbf p}_2 - {\mathbf p}_3\|\right).\eqno{(3.1)} $$ 
$$\langle \delta(0, 3)\rangle 
= \left\langle \|{\mathbf p}_1\| + \|{\mathbf p}_2\|+ \|{\mathbf p}_3\|-{1\over{\sqrt{3}}}\left(\|{\mathbf p}_1 - {\mathbf p}_2\|+\|{\mathbf p}_1 - {\mathbf p}_3\|+\|{\mathbf p}_2 - {\mathbf p}_3\|\right)\right\rangle.\eqno{(3.2)} $$
We note that $\delta(0, 3)$ itself is negative for the choice ${\mathbf p}_2 = -{\mathbf p}_1 \ne {\mathbf 0}$ and ${\mathbf p}_3 = {\mathbf 0}.$  However, this does not happen on the average. We have:

\heading{Theorem~1} $\langle \delta(0, 3)\rangle \geq 0.$\hfil

\heading{Proof}We know by boson symmetry that 
$$\langle\|{\mathbf p}_1\|\rangle = \langle\|{\mathbf p}_2\|\rangle =\langle\|{\mathbf p}_3\|\rangle:= k\eqno{(3.3)}$$
and
$$\langle\|{\mathbf p}_1-{\mathbf p}_2\|\rangle = \langle\|{\mathbf p}_1-{\mathbf p}_3\|\rangle =\langle\|{\mathbf p}_2-{\mathbf p}_2\|\rangle:= q.\eqno{(3.4)}$$
We may think of the $\{{\mathbf p}_i\}$, and their differences, as vectors, since they are used in momentum space where they become multiplicative operators. The six vectors in (3.1) are the six edges of a pyramid in $\Re^3;$  the norms, $\|{\mathbf p}_i\|$ and $\|{\mathbf p}_i-{\mathbf p}_j\|$, are the corresponding lengths of these six pyramid edges. The permutation symmetry of the wave function implies the equalities (3.3) and (3.4).  The mean difference  $\langle \delta(0, 3)\rangle$ is clearly smallest when the origin of the vectors $\{{\mathbf p}_i\}$ is at the centroid of the triangle formed by the differences $\{{\mathbf p}_i-{\mathbf p}_j\}.$ In this minimal case we know from elementary geometry that $q = \sqrt{3}k$; consequently, $\langle \delta(0, 3)\rangle = 0.$  It follows that in general $\langle \delta(0, 3)\rangle \geq 0.$ This completes the proof for the case $m=0,N = 3.$\qed
\section{Proof for the case $m= 0,~~ N= 4$.} 
We consider the case $N= 4$ and  $m = 0$ in (2.13). The six differences $\{{\mathbf p}_i - {\mathbf p}_j\}$ form a tetrahedron.  The average lengths  $q = \langle\|{\mathbf p}_i - {\mathbf p}_j\|\rangle$ are equal and force the tetrahedron to be regular.  Meanwhile, the four mean lengths
$k = \langle\|{\mathbf p}_i\|\rangle$ are again equal.  This symmetry occurs when the ${\mathbf p}$-origin is at the centroid of the tetrahedron, of, say, height $h.$ For such a tetrahedron we have 
$$h = \sqrt{\frac{2}{3}}q\quad{\rm and}\quad k = \sqrt{{3\over 8}}q.\eqno{(4.1)}$$
We may therefore write
$$\langle \delta(0, 4)\rangle 
= 4\left\langle\|{\mathbf p}_1\|\right\rangle -6\left({2\over 3}\right)\sqrt{{3\over 8}}\left\langle\|{\mathbf p}_1 - {\mathbf p}_2\|\right\rangle = 4k - 4\sqrt{{3\over 8}}q = 0.\eqno{(4.2)} $$

\nll Thus we have:

\heading{Theorem~2} $\langle \delta(0, 4)\rangle \geq 0.$\hfil
\section{Proof in the case $m \geq 0$,~$N = 3$} 
We consider
$$\displaylines{\delta(m, 3) 
= \left(\|{\mathbf p}_1\|^2 + m^2\right)^{\half}+\left(\|{\mathbf p}_2\|^2 + m^2\right)^{\half} + \left(\|{\mathbf p}_3\|^2 + m^2\right)^{\half}\hfill\cr
 ~~-(\third\|{\mathbf p}_1 - {\mathbf p}_2\|^2 + m^2)^{\half} - (\third\|{\mathbf p}_1 - {\mathbf p}_3\|^2 + m^2)^{\half} - (\third\|{\mathbf p}_2 - {\mathbf p}_3\|^2 + m^2)^{\half}\hfill{\rm(5.1)}}
$$ 
and
$$\displaylines{\langle\delta(m, 3)\rangle 
= \left\langle\left(\|{\mathbf p}_1\|^2 + m^2\right)^{\half}+\left(\|{\mathbf p}_2\|^2 + m^2\right)^{\half} + \left(\|{\mathbf p}_3\|^2 + m^2\right)^{\half}\right.\hfill\cr
 ~~-\left.(\third\|{\mathbf p}_1 - {\mathbf p}_2\|^2 + m^2)^{\half} - (\third\|{\mathbf p}_1 - {\mathbf p}_3\|^2 + m^2)^{\half} - (\third\|{\mathbf p}_2 - {\mathbf p}_3\|^2 + m^2)^{\half}\right\rangle.\hfill{\rm(5.2)}}
$$ 
\heading{Theorem~3}$\langle \delta(m, 3)\rangle \geq 0.$

\heading{Proof} The three vectors ${\mathbf p}_i$, $i = 1,2,3,$ and their three differences ${\mathbf p}_i-{\mathbf p}_j$ 
form six edges of a pyramid in $\Re^3;$  the norms, $\|{\mathbf p}_i\|$ and $\|{\mathbf p}_i-{\mathbf p}_j\|$, are the corresponding lengths of these six pyramid edges.  We now denote by $T$ the triangle formed by the three difference edges $\{\|{\mathbf p}_i-{\mathbf p}_j\|\}$. For convenience, we shall think of $T$ as lying in a horizontal plane and denote by $P$ the top vertex of the pyramid; without loss of generality, we shall speak of $P$ being above $T$.  We let $C$ be the point in the plane of $T$ vertically under $P$. We now pick the vertex of $T$ which contains ${\mathbf p}_1,$ and call this $V_1.$ In the plane of $T$ we construct a line from $V_1$ that is perpendicular to $CV_1$ and of length $m$, ending in the point $U_1$.  We then join $U_1$ to $P$ and observe that $\widehat{PV_1U_1} = \pi/2.$ Similar constructions are now made with the other two vertices $V_2$ and $V_3$ of $T$; the three line segments $U_iV_i$ are chosen to `flow' in the same way round the pyramid axis $CP$.  In fact, a new pyramid is formed by the three lines $PU_i$. By permutation symmetry we have that $\langle|PU_i|\rangle = k$ and $\langle|CU_i|\rangle = q$ where $i = 1,2,3,$ and moreover
$$\left\langle\left(\|{\mathbf p}_i\|^2 + m^2\right)^{\half}\right\rangle := k,\quad i = 1,2,3,\eqno{(5.3)}$$
and 
$$\left\langle\left({1\over 3}\|{\mathbf p}_i - {\mathbf p}_j\|^2 + m^2\right)^{\half}\right\rangle
:= q,\quad i,j = 1,2,3,\quad i\ne j.\eqno{(5.4)}$$
Since the position of $P$ which minimizes $k$ is $C,$ and symmetry is obtained on the average, we conclude by elementary geometry that $k \geq q.$
This inequality completes the proof of Theorem~3.\qed 
\section{Application to $N \geq 3$} 
For  $N\geq 3$ we can deduce a stronger lower bound than that provided by the $N/2$ bound, based on the result of Section~5.  If $E$ and $\Psi$ are the exact energy and corresponding wave function, we have $E = \left(\Psi, H\Psi\right)$ and therefore, by boson symmetry and Theorem~3, we have 
$$\eqalign{E &= {N\over 3}\left(\Psi,\left[(p_1^2 + m^2)^{\half} + (p_2^2 + m^2)^{\half}+ (p_3^2 + m^2)^{\half}\right.\right.\cr
&~~~~~~~+ \left.\left.{{N-1}\over 2}\left(V(r_{12})+V(r_{13})+V(r_{23})\right)\right]\Psi\right)\cr
&\geq N\left(\Psi,\left[({1\over 3}p_{12}^2 + m^2)^{\half} +{{N-1}\over 2}V(r_{12})\right]\Psi\right)\cr
&\geq N\left(\Psi,\left[({4\over 3}p^2 + m^2)^{\half} +{{N-1}\over 2}V(r)\right]\Psi\right),}$$
where ${\mathbf r} = {\mathbf r}_1-{\mathbf r}_2$  and ${\mathbf p} = \half({\mathbf p}_1-{\mathbf p}_2) = {\mathbf p}_{12}$.  Thus we have, for $N\geq 3$, $m \ge 0,$ and $\|\psi(r)\| = 1$:

\heading{Theorem~4}

$$E \geq E^L_{N/3} = N\inf_{\psi}\left(\psi, \left[\left({{4}\over{3}}p^2 + m^2\right)^{\half} + {{N-1}\over{2}}V(r)\right]\psi\right).\eqno{(6.1)}$$

In similar fashion, we can relate the $N$-body problem for $N\ge4$ and $m = 0$ to a reduced $4$-body problem based on Theorem~2.  Specifically, we have for $N \ge 4$, $m = 0,$ and $\|\psi(r)\| = 1$:

\heading{Theorem~5}

$$E \geq E^L_{N/4} = N\inf_{\psi}\left(\psi, \left[\left({{3}\over{2}}\right)^{1\over 2}\|{\mathbf p}\| + {{N-1}\over{2}}V(r)\right]\psi\right).\eqno{(6.2)}$$

Theorems~4 and 5 summarize the principal results of this paper.
\section{The linear potential $V(r) = r$ with $m=0$} 
The lower bounds we have found all presume that the bottom of the spectrum of a scaled one-body problem can be found.  For Salpeter Hamiltonians, this task itself may not be trivially easy, although more tractable than for the many-body problem.  For the operator $H = \|{\mathbf p}\| + r $ in three dimensions, we have at our disposal the accurate value $e = 2.2322$, for example, from the work of Boukraa and Basdevant \cite{Boukraa89} (the linear potential has also been considered by Pirner and Wachs \cite{PirnerWachs97} in an application to quark systems). By elementary scaling arguments we therefore have for the one-body problem    
$$H = a p + b r\quad\rightarrow\quad E(a, b) = (a b)^{\half}E(1,1) = (a b)^{\half}e,~~a, b >0,~~e = 2.2322.\eqno{(7.1)}$$
This relation will generate all the energy lower bounds for $N$-body problems with this potential.  We shall use the notation $E^L_{N/2},$ $E^L_{N/3},$ and $E^L_{N/4},$ for the lower bounds given by equations (1.3), (6.1), and (6.2), and $E_c$ for the conjectured bound (2.11).  The formula (7.1) then allows us to derive formulas for these energies.  Explicitly we find:
$$E^L_{N/2} = N\left(\frac{N-1}{2}\right)^{\half}e,\ N\ge 2\eqno{(7.2a)}$$
$$E^L_{N/3} = N\left(\frac{N-1}{\sqrt{3}}\right)^{\half}e,\ N\ge 3\eqno{(7.2b)}$$
$$E^L_{N/4} = N\left(\frac{3(N-1)^2}{8}\right)^{1\over 4}e,\ N\ge 4\eqno{(7.2c)}$$
$$E_c^L =  N\left(\frac{(N-1)^3}{2N}\right)^{1\over 4}e,\ N\ge 2.\eqno{(7.2d)}$$

In order to find an upper bound, we follow Ref.~\cite{Hall04}~and use a Gaussian wave function, which
we write initially in the form
$$\Phi(\rho_2,\rho_3,\dots,\rho_N)
=C\exp\left(-\frac{1}{2}\sum_{i=2}^N\rho_i^2\right)= \prod\limits_{i = 2}^{N}\phi(\rho_i),\quad C = \left(\frac{2}{\pi^{\frac{1}{4}}}\right)^{N-1},\eqno{(7.3)}$$
where the constant $C$ is chosen to ensure the normalization of each radial factor $\phi$ on $L^2([0,\infty),r^2 dr).$
The boson symmetry of the trial function allows us to write $E \leq E_g^U = \left(\Phi,H\Phi\right),$ where we have
$$E_g^U =  \left(\Phi,\left[N\|{\mathbf p}_N\| + \gamma V(\|{\mathbf r}_1-{\mathbf r}_2\|)\right]\Phi\right).\eqno{(7.4)}$$
The identity (2.4) and the lemma proved in \cite{Hall03} (which allows us to remove the operator term $\pi_1$) imply
$$E_g^U =  \left(\Phi,N\sqrt{\frac{N-1}{N}}\|\pi_N\| + \gamma V(\sqrt{2}\rho_2)\Phi\right).\eqno{(7.5)}$$
The permutation symmetry of the Gaussian function in the {\it relative} coordinates and the factoring property allow us to replace 
$\pi_N$ by $\pi_2 \equiv \sqrt{2}{\mathbf p}.$  We write the conjugate variable to ${\mathbf p}$ as ${\mathbf r}\equiv\sqrt{2}\rho_2$, so that $V(r) = r$, and the wave function becomes $\phi(r).$
By introducing an additional scale parameter $\sigma > 0,$ we then find
$$E_g^U = N\left(\sqrt{\frac{2(N-1)}{N}}\frac{1}{\sigma}\langle p\rangle + \frac{N-1}{2}\sigma\langle r\rangle\right).\eqno{(7.6)}$$ 
 Since the Gaussian radial function $\phi(r)$ is form invariant under the 3-dimensional Fourier transformation, we have the equality
$$\langle p\rangle = \langle r\rangle = \frac{2}{\sqrt{\pi}}.$$
Since the minimum of the form $\alpha /\sigma + \beta\sigma$ over the scale $\sigma >0 $ is $2(\alpha\beta)^{\half},$ we arrive at the following explicit formula for the Gaussian upper bound: 
$$E_g^U = 4N\left(\frac{(N-1)^3}{2N\pi^2}\right)^{\frac{1}{4}}, \ N \ge 2.\eqno{(7.7)}$$
We can immediately test this formula for the case $N = 2$ to obtain $E_g^U = 3.19154,$  which is to be compared with the accurate numerical value $E = \sqrt{2}e = 3.1568.$ More generally, we exhibit in Table~1 ratios $R_{X} = E_g^U/E^L_{X},$ where $X$ is $N/2, N/3, N/4$ or, for the conjectured lower bound, $R_c = E_g^U/E_c^L.$  The percentage error in the determination of the energy by the bounds is approximately $50(R-1)\%.$  The monotonic behaviour of $R$ with $N$ follows from the `distance' of $N$ from the size of the sub-system whose lower bound is best possible; if the conjecture were true, the quality of the lower bound would be the same for all $N$.\medskip

\fulltable{\label{table:table1}Ratios of upper to lower energy bounds $R_X = E_g^U/E_X^L,$ where $X = N/2,N/3, N/4;$ the ratio for the conjectured lower bound is $R_c = E_g^U/E_c^L.$\\ }
\begin{tabular}{|c|c|c|c|c|c|c|c|}
\hline
& $N=2$ & $N=3$ & $N=4$ & $N=5$ & $N=6$ & $N=10$ &$ N\rightarrow\infty$\\ \hline
$R_{N/2}$&1.011 & $1.08639$ & $1.11886$ & $1.13706$ & $1.14872$ & $1.17104$ & $1.20229$ \\ \hline
$R_{N/3}$& & $1.011$ & $1.04121$ & $1.05815$ & $1.069$ & $1.08977$ & $1.11886$ \\ \hline
$R_{N/4}$& &  & $1.011$ & $1.02745$ & $1.03799$ & $1.05815$ & $1.08639$ \\ \hline
$R_{c}$&1.011 & $1.011$ & $1.011$ & $1.011$ & $1.011$ & $1.011$ & $1.011$ \\ \hline
\end{tabular}
\endfulltable

\section{Conclusion}
If a system of $N$ identical particles is bound together by attractive pair potentials, the Hamiltonian $H$ has $N$ kinetic-energy terms and $\gamma = {N\choose2}$ potential terms. If the kinetic energy of the centre-of-mass can be subtracted off, then the number of kinetic-energy terms is reduced by one, and we would expect to obtain an expression of the form $E = \langle H\rangle = \langle (N-1)K + \gamma V\rangle.$  The $N$-body energy $E$ is then bounded below by the lowest energy ${\mathcal E}$ of a `reduced' one-body operator of the form ${\mathcal H} = (N-1)K + \gamma V;$ if the boson-symmetry requirement of the $N$-body wave function is not too stringent, then this lower bound is at the same time a good approximation.  This story is realized exactly for the nonrelativistic problem \cite{Hall67}: for the special case of the harmonic oscillator, ${\mathcal E}$ yields the exact energy $E$ of the many-body system.  The reduction details depend on the quadratic form of the nonrelativistic many-body kinetic-energy operator and the identities (2.3) for quadratic expressions in Jacobi relative coordinates. 

For the semirelativistic counterpart, one generally loses the quadratic form in $H$ and, along with it, the immediate reduction.  An alternative reduction to the $H_{N/2}$ Hamiltonian is always possible and is important theoretically, but the resultant lower energy bound is weak.  A quadratic form is returned to the potential in $H$ in the special case of the harmonic oscillator, and this yields \cite{Hall04} a very sharp bound on the energy, though not now the exact solution, except in the Schr\"odinger limit $m\rightarrow \infty.$  For general pair potentials, we have constructed a new Hamiltonian $H_c$ that is translation invariant, both in coordinate and momentum space, and which reduces to the usual two-body Hamiltonian for $N=2$, and generally to $Nm+H_S$ in the large-$m$ limit, where $H_S$ is the Schr\"odinger Hamiltonian with the centre-of-mass kinetic energy removed.  A reduction $\langle H_c\rangle =\langle {\mathcal H}\rangle \geq {\mathcal E}$ of $H_c$ to a one-body Hamiltonian ${\mathcal H}$ immediately follows.  This is useful for the study of the many-body Hamiltonian $H$ whenever it can also be established that $\langle H\rangle \ge \langle H_c\rangle.$  We conjecture that this is always true.  In the present paper we have proved the conjecture for $N = 3,$ and for $N= 4$ if $m = 0;$ it is also true for the harmonic oscillator, and generally in the large-$m$ limit.   For the case of static gravity $V(r) = -1/r,$  the conjecture yields the identical result to the energy bound we have established by a completely different argument, valid specially for this potential \cite{Hall06}.

 \section*{Acknowledgement}
One of us (RLH) gratefully acknowledges both partial
financial support of his research under Grant No.~GP3438 from~the Natural
Sciences and Engineering Research Council of Canada and hospitality of the
Institute for High Energy Physics of the Austrian Academy of Sciences in
Vienna.
\medskip

\end{document}